\documentclass[preprint]{aastex}

\usepackage{lineno,hyperref}

\begin{document}


\title{Gridlock Models with the IBM Mega Traffic Simulator: Dependency on Vehicle Acceleration and
Road Structure}

\author{Bruce G. Elmegreen\footnote{belmegreen@gmail.com}}
\author{Tayfun Gokmen}
\author{Biruk Habtemariam}
\affil{IBM Research Division, T.J. Watson Research Center, Yorktown Heights, NY
10598}

\begin{abstract}
Rush hour and sustained traffic flows in eight cities are studied using the IBM Mega
Traffic Simulator to understand the importance of road structures and vehicle
acceleration in the prevention of gridlock. Individual cars among the tens of
thousands launched are monitored at every simulation time step using live streaming
data transfer from the simulation software to analysis software on another computer.
A measure of gridlock is the fraction of cars moving at less than 30\% of their
local road speed. Plots of this fraction versus the instantaneous number of cars on
the road show hysteresis during rush hour simulations, indicating that it can take
twice as long to unravel clogged roads as fill them. The area under the hysteresis
loop is used as a measure of gridlock to compare different cities normalized to the
same central areas. The differences between cities, combined with differences
between idealized models using square or triangular road grids, indicate that
gridlock tends to occur most when there are a small number of long roads that
channel large fractions of traffic.  These long roads help light traffic flow but
they make heavy flows worse. Increasing the speed on these long roads makes gridlock
even worse in heavy conditions. City throughput rates are also modeled using a
smooth ramp up to a constant vehicle launch rate. Models with increasing
acceleration for the same road speeds show clear improvements in city traffic flow
as a result of faster interactions at intersections and merging points. However,
these improvements are relatively small when the gridlock is caused by long roads
having many cars waiting to exit at the same intersection. In general, gridlock in
our models begins at intersections regardless of the available road space in the
network.
\end{abstract}




\section{Introduction}

Traffic flow in cities differs from traffic on highways because city driving has
many types of streets with various road speeds, and it has frequent intersections
where driver judgement, signs, and lights determine the right of way. Cities also
have complex road networks that can make them difficult to model. Here we use the
IBM Mega Traffic Simulator to simulate rush hour traffic and sustained flows in
eight cities. The purpose is to understand the influence of driver acceleration and
road structure on the development and dissipation of gridlock, a condition where a
high fraction of drivers are unable to move at the normal road speed because of
congestion.  Reviews of traffic models are in \cite{newell65,helbing01,bellomo11}.

A discriminant for traffic mobility at merge points is the dimensionless parameter
$aD/v^2$ for vehicle acceleration $a$, separation $D$ and speed $v$. For example,
when this number equals 1, cars can accelerate from a stop and fit between two cars
moving at speed $v$ and separation $D$. High values corresponds to easy merging
conditions from a stop, while low values make merging difficult because there is no
room to fit without precise timing.  For any given city, the average separation $D$
scales inversely with the ratio of the number of cars on the road, $N$, to the total
occupied road length $L$, which is the sum of the product of all the utilized road
lengths and their corresponding number of lanes. Then $aL/(Nv^2)$ might be
considered an important, analogous, quantity that should be as high as possible. Low
values occur during rush hour when $N$ is high. They also occur when traffic tends
to prefer a few main boulevards and cross streets, decreasing $L$ for fixed $N$.
Slow speeds help, as do high accelerations. However, the average road length per
car, $L/N$, is not as decisive an indicator of potential flow problems as the local
separation $D$ if the congestion tends to occur near an intersection with free flow
on the road before that. The best indicators are the most local and instantaneous,
which makes traffic analysis extremely data rich on the scale of whole cities.

High acceleration also relieves congestion on roads without intersections, such as
highways.  For a discussion of highway flow, oscillation instabilities, and possible
solutions, see, for example, \cite{daganzo96,li02,baran13,li14}.  If the mean flow
rate per lane, $Nv/L=v/D$ measured in cars per second, exceeds the outflow rate at
the rarefraction front leading a pack that forms, which is $(0.5a/D)^{0.5}$, then
more cars will join the pack at the back end than can leave it at the front end. The
pack therefore grows. This outflow rate is the inverse time that it takes to
accelerate up to the point where the vehicle separation is $D$. The pack does not
grow when $aD/v^2$ exceeds 2. Cars that accelerate more quickly and roads that have
lower speeds each lead to more stable conditions.

Car acceleration is useful to consider as a variable in models of traffic control.
Future communication systems between cars could aid in the control of this variable
by prompting the driver to adjust the car's acceleration -- up to a reasonable limit
-- when needed to improve the flow \cite{ge13}. Autonomously driving cars could also
have optimum accelerations. As it is now, most drivers accelerate at fairly low
rates in city conditions. For example, acceleration up to 30 miles per hour in half
of a city block, which is $\sim0.05$ of a mile, is the equivalent of only 11\% of
the acceleration of gravity, $0.11g$.  The fastest sports car accelerates from 0 to
60 at about $1g$ \cite{anderson09}. The maximum deceleration during braking is also
about $1g$ with a typical value of unity for the coefficient of rolling friction
between rubber tires and the road.

In city conditions, when cars cannot merge after turning from one road to another at
an intersection, traffic backs up. We would like to model such gridlock and see if
it can be relieved by increasing the key dimensionless quantity discussed above,
$aD/v^2$. We do this using the IBM Mega Traffic Simulator code
\citep{osogami12,osogami13} applied to 8 cities using road networks and speeds from
www.openstreetmap.com, and using vehicle launch rates that simulate either a rush
hour, which is done with a Gaussian launch rate profile, or a steady flow, which is
done with a half-Gaussian ramp up to a steady flow.  We randomize the origin and
destination positions for this rate profile, and then the code chooses the route in
advance. Idealized road networks are studied also. Several measures of gridlock are
employed and tested against variations in the launch rate and acceleration. Certain
cities are found to be consistently worse than others depending primarily of the
number of difficult intersections.

In Section \ref{method}, overviews of the IBM Mega Traffic Simulator and the traffic
model are given. We also discuss a method to stream results from the Simulator into
an analytics program on another computer. Section \ref{rush} shows the rush hour
results comparing 8 cities in a standard model, and Section \ref{ideal} considers
idealized cities. Section \ref{accel} shows results for these cases again with
higher accelerations. Section \ref{steady} discusses steady flow-through models for
the 8 cities. The conclusions are in Section \ref{conclusions}

\section{Method}
\label{method}

\subsection{IBM Mega Traffic Simulator}

The IBM Mega Traffic Simulator, called {\it Megaffic} in what follows, is an
agent-based traffic simulator \citep{osogami13} that uses street maps from
http://www.openstreetmap.com and accepts as input a table of origin and destination
points on a rectangular grid for each second of time. Streets are designated as
``primary,'' ``secondary,'' and so on, with different speeds for each type ranging
from 80 km hr$^{-1}$ for ``motorways'' to 20 km hr$^{-1}$ for ``residential.'' The
algorithm to determine the route for each car from the origin and destination grid
points is discussed in \cite{imamichi13}. These routes follow from the
origin-destination table and are fixed for each car at the start of the simulation.
Driver preferences with regard to travel time, distance, and number of turns are
considered.

The program uses the Gipps model for driver action \cite{gipps81}, which contains an
acceleration value $a$, evaluated here with a Gaussian probability distribution
function with standard deviation $\sigma_{\rm a}$. Our nominal value is
$a=1.7\pm0.3$ m s$^{-2}$ but different values are used for experimentation. For
reference, the vehicle lengths are all assumed to be 4.46 meters, the time step is 1
second, and the reaction time in the Gipps model is $2/3$ second.  The program also
uses the lane-selection model in \cite{toledo03}.

Although {\it Megaffic} is highly sophisticated as a simulation tool, it is still
under development. It moves cars according to standard models along realistic road
networks but there are elements of real traffic flow that are not present yet. For
example, the models used here do not have traffic lights at intersections, nor can
drivers change their routes to respond to changing road congestion. Still, it is
useful as a comparison between cities and to test some basic properties of traffic
flow in complex networks with tunable conditions, such as driver acceleration.

Use of {\it Megaffic} also allows monitoring of every car at every second, something
that is not possible in real cities. The data rates are enormous for this, however.
Traffic flow is an interesting problem from the point of view of data volume. Every
car among tens or hundreds of thousands of cars is doing something interesting every
second, such as braking, accelerating, turning, or interacting with other cars that
are only seconds away in time.  Understanding the source, origin and control of
potential bottlenecks requires second by second monitoring at key locations, and for
some cities, at many locations. Thus the problem has a large dynamic range in both
space and time dimensions. For example, the range of spatial scales is the ratio of
the city size to the car size (e.g., $\sim2000$-squared), or the total road length
to the car size ($\sim 50000$), while the range of time scales may be determined by
1 second intervals for an hour or two ($\sim5000$). This dynamic range even for
inner city areas can exceed a billion distinct information elements during rush
hour. For a large city and with commuters from the suburbs, the information can
exceed a trillion elements.

\subsection{Streaming Analytics}

{\it Megaffic} enables traffic analysis on microscopic levels while generating
massive amounts of simulation data. At each simulation time step, each car's
longitude, latitude, speed, acceleration, distance to the leading car, road of
travel, CO$_2$ output, and other quantities, are computed and updated for the next
time step. For a small city model, these values can be written to storage for later
use, but for a large city with many cars traveling a long time, the data volume can
be too large to store, and only the time-averaged or integrated quantities can be
saved. This inability to write what is essentially every variable at every time step
is common for computer simulations, which typically store only values at widely
spaced intervals to limit the total data volume.  Traffic flow is an intrinsically
data-rich problem, however, where something interesting and important has the
potential to occur at every time step for every car. Other physical problems are
like that too, such as turbulence, weather forecasting, and financial markets.

For {\it Megaffic}, a lack of transparency to the state of the simulation at every
time step makes it very difficult to perform a car-by-car analyses and
visualizations that represents the real experiences of drivers in a large city.
Also, the usual procedure of writing to disk during the simulation and they
analyzing the results later only gives visibility to the problem after it is too
late to change anything.

In order to overcome these limitations, we added a streaming capability
to the \textit{Megaffic} software. At each simulation time step,
each car state was packaged into a message and streamed to another process running
on another computer which is only responsible for analyzing and visualizing the
simulated data as it arrives. This approach decouples the simulation program from
the analysis program. The analysis can then handle arbitrarily large amounts of data
without ever requiring it to be stored. Streaming analytics also makes it easier to
modify the analysis method, seeking out unexpected features, for example, without
altering the simulation code while it is running.

In our application, the communication between processes used UDP protocol as it does
not require hand-shaking and connection between them. This approach guarantees that
the simulation program can still run at a full speed without worrying about the
latency of the network or the analysis program on a different computer. However,
since UDP protocol does not guarantee delivery and ordering, the analysis program
needs to compensate for or be impervious to occasional transmission errors.

We used {\it MATLAB} and IBM's {\it Infosphere Streams} as two examples for the
analysis software. In some cases we ran {\it Megaffic} on a multinode IBM Cloud
computer and streamed the data to a socket on a desktop computer, where it was
retrieved on-the-fly and put through {\it MATLAB} or {\it Infosphere Streams}. In
other cases we ran {\it Megaffic} on one desktop computer and streamed the results
to another running these programs.  In all cases, we were able to visualize and
monitor the traffic state at single-second time resolution and on an individual car
basis. Since the streaming was performed while the simulation was in progress, a
real time display of every car on the road was realized.

Streaming was also used to display and map instantaneous gridlock measures (Section
\ref{rush}). By viewing where and when the slow spots occurred on a city road map,
the positions and speeds of other cars around them, the level of congestion on
adjacent roads where the slow cars needed to merge, the relative speeds of cars on
the adjacent roads, the road structures, and so on, we could watch the gridlock
patterns develop and understand their origins, such as the difficulty of merging
onto new roads at certain intersections. We could also try various fixes in
different {\it Megaffic} simulations, such as higher accelerations for all cars, and
understand quickly how well they worked by watching the same cars at the same
intersections when the problem was solved.

One study, for example, considered the role of a few vehicles with low
accelerations. We noted that these vehicles did not affect the overall city
congestion much and wondered why. So we tagged them and watched them move through
the city streets along with all of the other cars. The laggards accumulated lines of
other cars behind them between intersections, as expected, but as soon one left an
intersection, the cars trapped behind it dodged off to other roads at their normal
accelerations and the line temporarily went away.  This was a different behavior
compared to the long lines that accumulated at permanently bad intersections, which
were the most common cause of gridlock.

{\it InfoSphere Streams} was developed to ingest and analyze information in large
data streams to enable on-the-fly big data applications \citep{infosphere}. It
provides built-in operators for basic streaming operations, and has a Stream
Programming Language (SPL) where end-users can create their own operators. For this
project, we used the built-in \textsc{udp sink} operator to ingest and convert the
aggregated message stream from \textit{Megaffic} into a flow of tuples, and the
built-in \textsc{split} operator to extract a single car state from a tuple. Then we
defined an operator to calculate our gridlock condition, i.e., the fraction of cars
moving below $30\%$ of their corresponding road speed (Section \ref{rush}), and
other interesting quantities. For visualization, the \textsc{http tuple view}
operator was used to stream the car position to a display program on the internet
while \textit{Megaffic} was running. Although the visualization tools were limited
for InfoSphere Streams, this software is a much more scalable option than MATLAB for
analyzing large amounts of data.

\section{Launching Rates for Cars}

\begin{figure}
\includegraphics[width=0.9\textwidth]{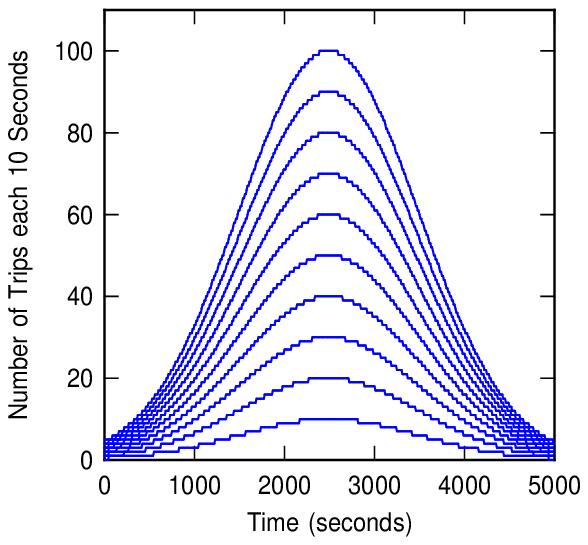}
\caption{Trip launch rates for rush hour simulations. Rates are in cars or trips
per 10 seconds. } \label{counttrips_1c10s500cy}\end{figure}

To simulate rush hour, we launched cars in various cities using a 10x10 grid inside
the central 10 km by 10 km square road network. Normalization of each city to the
same area mitigates trivial scaling differences when we compare the results. The
cars were launched in batches, with some number $R$ at a time using randomly chosen
origin and destination points. The launching times were separated by 10 seconds to
space them out along the adjacent streets. Thus the launch rate was $R/10$ cars per
second within the 100 km$^2$ area. To simulate a rush hour, we set
\begin{equation}
R(t)=R_0\exp\left(-0.5\left[t-t_0\right]^2/\sigma^2\right)
\label{lr}
\end{equation}
where $t=0$, 10, 20, 30, ... seconds up to some maximum time $t_{\rm max}$, taken to
be 5000 seconds in many cases but varied to study the impact of spacing out cars
during rush hour. To make a smooth Gaussian launch pattern, we took $t_0=t_{\rm
max}/2$ and $\sigma=t_{\rm max}/5$. Sample launch patterns are shown in Figure
\ref{counttrips_1c10s500cy}. The curves are boxy because the launch numbers have to
be integers for discrete cars. Rush hour simulations using launch rates like this
are discussed in section \ref{rush}.

In another set of experiments, cars were launched at a rate that has a half-Gaussian
ramp with $\sigma=1000$ seconds up to the peak at 2500 seconds, and then remains
constant thereafter for at least 20,000 seconds. After the initial ramp up, this
model simulates a steady flow of cars to determine what a city can sustain without
gridlock conditions. These steady flow models are discussed in section \ref{steady}.

\section{Results for Rush Hour Simulations}
\label{rush}

Cars launched into the road network of a city accelerate up to the road speed and
move around, negotiating other cars with a no-collision rule and turning from one
street to another according to pre-determined routes. Cars that hesitate or stop at
intersections and other places cause the cars behind them to slow down or stop as
well, as in a normal traffic flow.

We are interested in finding a good diagnostic for gridlock conditions, when a high
fraction of cars cannot move at the nominal road speed. To search for such a
diagnostic, we tried various things, such a the distribution function of trailing
distances between cars, the fraction of cars stopping, and so on. The clearest
diagnostic we found was the fraction of cars with an instantaneous speed below some
fraction of the nominal road speed. To find this limiting fraction, we plotted
histograms of the ratio of car speed to local road speed in fixed intervals of time.

\begin{figure}
\includegraphics[width=1.\textwidth]{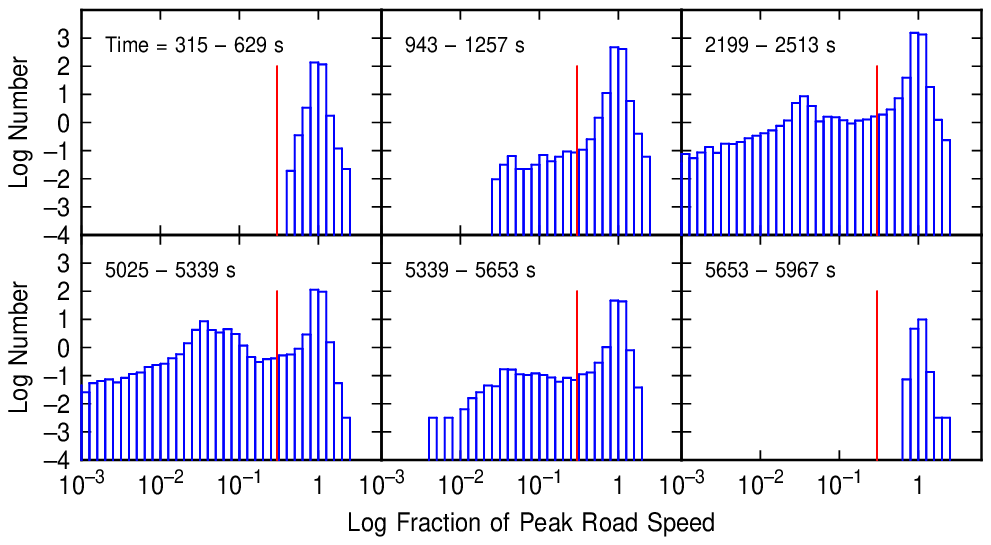}
\caption{Distributions of the ratio of instantaneous car speed to road speed for several
intervals of time in a rush hour simulation of the central 10 km square region in
Washington DC. At early times, there are very few cars and none are moving slower than
30\% of the road speed (vertical red lines). As more cars enter the roads, the
total number increases and the fraction of cars that move slowly increases too.
The total duration of the rush hour start times is as shown in figure 1, 5000 seconds,
but histograms here in the lower panel show significant numbers of cars and high
fractions that are moving slowly long after this time. The time at the peak launch
rate, 2500 seconds, corresponds to the upper right panel. This delay in clearing
of the congestion, compared to the relatively fast time for it to
build up, corresponds to an asymmetry of traffic flow, leading to the
hysteresis shown in Figure 3.} \label{his_speedtime_log_washington_80_500_short}\end{figure}

Figure \ref{his_speedtime_log_washington_80_500_short} shows an example with 6 equal
time intervals out of the total travel time of 5973 seconds for a rush hour
simulation in Washington DC. The peak launch rate in equation \ref{lr} is $R_0=80$
cars per 10 second interval and the launch window is $t_{\rm max}=5000$ seconds.
Each plotted interval represents 1/19th of the total time when cars are on the road.
The other time intervals between these look about the same. The total time is longer
than $t_{\rm max}$ because cars continue to move to their destinations after the
last one is launched.

Figure \ref{his_speedtime_log_washington_80_500_short} shows that at the beginning
of the simulation, in the time interval from 315 to 629 seconds (top left), most
cars are within 30\% of the local road speed, whatever that is (the local road speed
varies from car to car, depending on which type of road that car is on). There is no
significant congestion. At time progresses, more and more cars dip below 30\% of
their road speed, which is indicated by the red vertical line. At the time of peak
launch rate, which is $t_0=2500$ seconds, a high fraction of cars are moving below
30\% of the local road speed, and gridlock prevails (there are even more cars not
plotted and out of range to the left in the figures, moving slower than 0.1\% of the
road speed). This bad condition continues until well after the last car is launched,
with a significant fraction moving slowly even at $t=5339-5653$ seconds. Only after
$\sim5600$ seconds do the roads clear up.

The fraction of cars moving slower than 30\% of the local road speed is considered
here to be a good measure of bad driving conditions after experiments like this.
There is hysteresis in the congestion, with bad conditions asymmetrically shifted
toward late times compared to the time of peak launch. Figure
\ref{his_speedtime_log_washington_80_500_short} shows that roads fill up quickly but
drain slowly.

\begin{figure}
\includegraphics[width=.8\textwidth]{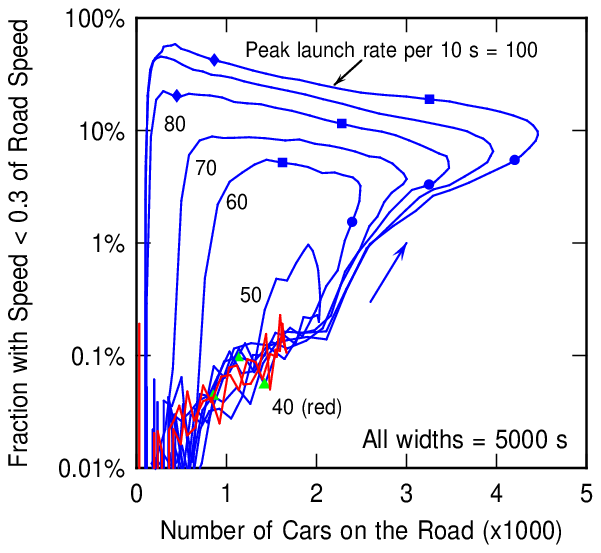}
\caption{The fraction of the cars moving slower than 30\% of their local
road speed is shown as a function of the number of cars on the road
for rush hour simulations in Washington DC. Each curve has a different
peak launch rate, $R_0$, with lower $R_0$ corresponding to less
congestion because the number of cars on the road at any one time is lower.
The curves are traversed in a counter clockwise direction, as shown by the
arrows. The green triangles in the lower part of the ascending curves
corresponds to a time midway up the rising part of the rush hour model,
i.e., at $t_0/2=1250$ seconds. The dots correspond to the time of peak
launch rate, 2500 s, the squares are at 3750 s, and the diamonds are at
5000 seconds, when cars stop entering the road. For high launch rates,
the worst traffic jams occur after the cars stop entering the road (the peak
in the curves is counter clockwise from the diamonds) because
the last tens of percent of cars have no where to go. } \label{fractspeedtime_rush}\end{figure}

The main results in Figure \ref{his_speedtime_log_washington_80_500_short} are made
more concise by plotting the fraction of cars moving slower than 30\% of the road
speed versus the number of cars currently on the road. Such plots are shown in
Figure \ref{fractspeedtime_rush} for Washington DC with seven different peak launch
rates, $R_0=40$, 50, 60, ..., 100 cars per 10 seconds, all with $t_{\rm max}=5000$
seconds. Each curve is a single experiment of a complete rush hour in the road
network using the same 10x10 grid for origin and destination points, although all of
these points are random and different for each case. The resulting curves do not
depend on these random routes significantly for a given launch rate. The wiggles in
the curves reflect the details of individual cars stopping and starting. As time
progresses, the position of a simulation on a curve moves counter clockwise, as
indicated by the blue arrow. Alternate curves have fiducial markers indicating the
time: green triangles are at $t_{\rm max}/4$ (i.e., 1250 seconds; these are
difficult to see as they occur in the lower noisy part of each curve); filled
circles are at $t_{\rm max}/2$, squares are at $3t_{\rm max}/4$, and diamonds are at
$t_{\rm max}$. The worst gridlock occurs at the top of each curve, where the
fraction of cars moving slower than 30\% of the road speed is high, often exceeding
10\%. The corresponding time is between 30\% and 100\% of the maximum launch time.
As expected, the gridlock improves and the number of cars on the road decreases as
the peak launch rate decreases.

\begin{figure}
\includegraphics[width=.8\textwidth]{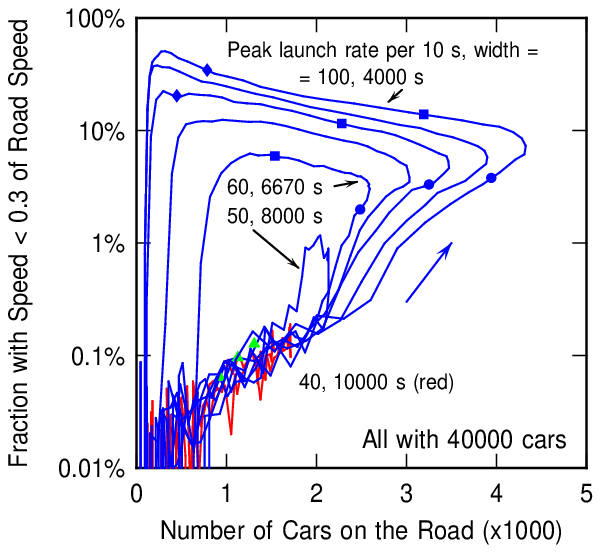}
\caption{The fraction of slow cars is plotted versus the number of cars
on the road for Washington DC in 7 cases with different total time spans for
the rush hour, all with the same total number of cars launched.
As the rush hour is spread out in time, the fraction of slow cars
decreases. Even small changes in the duration of rush hour lead to
large changes in the slow fraction, considering the ordinate is in
logarithmic coordinates. The pair of numbers indicated for each
curve is the launch function pair, $(R_0,t_{\rm max})$,
in the notation of equation \ref{lr}.} \label{fractspeedtime_rush_varywidth}\end{figure}

Figure \ref{fractspeedtime_rush_varywidth} shows the well-known result that gridlock
improves if the rush hour time is prolonged for the same total number of cars. The
different curves have different $R_0$ and $t_{\rm max}$ with a constant product
$R_0t_{\rm max}$, which is proportional to the total number of cars launched. As
$R_0$ decreases and $t_{\rm max}$ increases, the peak fraction of slow cars
decreases. The plot has a logarithmic ordinate, so the decrease is rapid with small
increases in $t_{\rm max}$. Doubling the maximum time from 4000 s to 8000 s changes
the flow from 20\%--50\% gridlocked to less than 1\% gridlocked.

Now we consider eight different cities, all with the inner 10 km square used for the
road network. These cities span a variety of network shapes, from regular grids, as
in Indianapolis and Beijing, to highly convoluted small streets, as in London,
Istanbul, and Damascus. Some have large waterways running through them with several
bridges going from one side of the city to the other (e.g., Washington D.C.,
Istanbul).

\begin{figure}
\includegraphics[width=.8\textwidth]{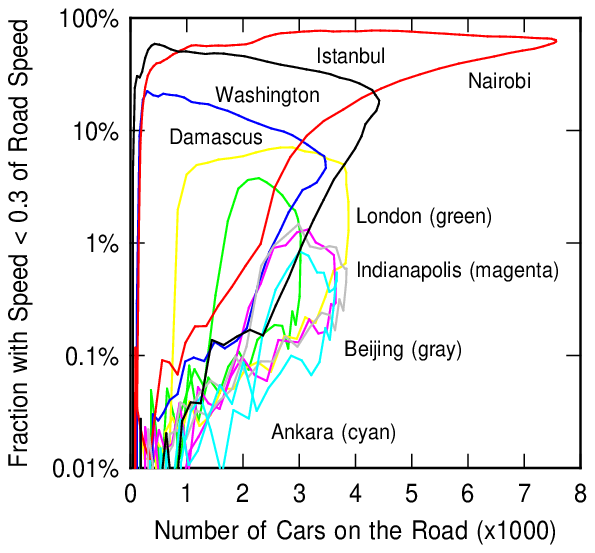}
\caption{The fraction of slow cars is plotted versus the number of cars
on the road for 8 cities with various road types. All of the curves are for
the same rush hour model with $R_0=80$ cars per 10 seconds, and $t_{\rm max}=5000$
seconds. Some cities get congested much more easily than others, as shown by the
high values of the instantaneous slow fraction. Time increases counter clock wise in
each loop. The correspondence between color and city is preserved in the next two
figures, for clarity.} \label{fractspeedtime_rush_allcities}\end{figure}

Figure \ref{fractspeedtime_rush_allcities} shows the results for the 8 cities. They
all have the same launch rate and duration, $R_0=80$ cars per 10 s, $t_{\rm
max}=5000$ s. The cities with high looping curves in the figure are more easily
congested in our models than the others. Note that the total number of cars launched
is the integral under the launch rate in equation 1, $(2\pi)^{0.5}\sigma
R_0/10=20053$.  The maximum number on the road at any one time for most of the
cities is about 20\% of this integral, which indicates that most cars get where they
are going even when there is severe gridlock elsewhere. Nairobi roads reach a peak
count of 7567 cars and a peak fraction of cars slower than 30\% of the road speed
equal to 77.4\%. Four cities have slow-car fractions of about a per cent or less.

What differences between cities contribute to a range of slow-car fractions even
when they have same launch rates and city areas? We considered that the differences
could be the total road capacity for all of the occupied roads, or perhaps the
normalized capacity which is the road length per car, or the average number of cars
per road, or perhaps the average road speed per occupied road. These quantities were
measured at every second and a representative sample for Figure
\ref{speedlength_80_500dots} was taken at two specific time steps, the time of peak
launch, $t_0=2500$ s, and midway down the Gaussian after the peak, at $t=3750$ s.
The abscissa in the plots is the integral under the hysteresis loop in Figure
{\ref{fractspeedtime_rush_allcities}, which is a measure of gridlock. Filled circles
are for the first time, and squares are the second time (the same symbols as in
Figures \ref{fractspeedtime_rush} and \ref{fractspeedtime_rush_varywidth}). Colors
represent cities as in Figure \ref{fractspeedtime_rush_allcities}.

\begin{figure}
\includegraphics[width=.8\textwidth]{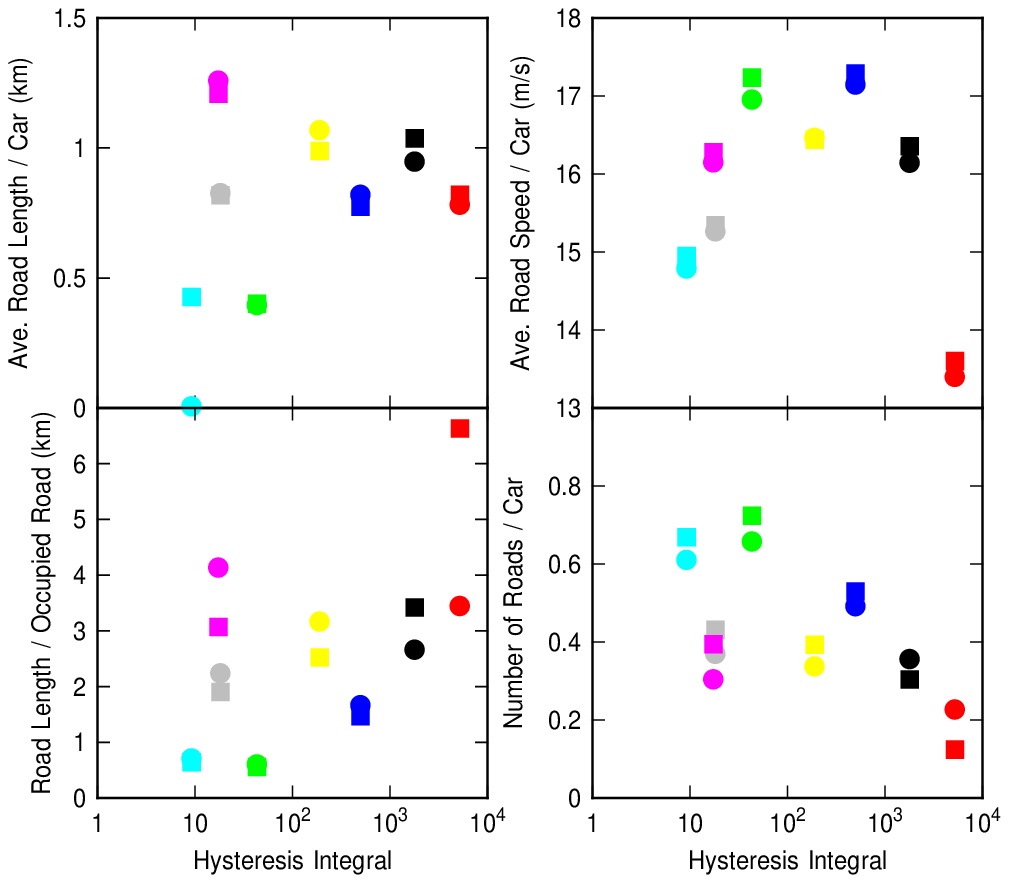}
\caption{Distributions of various quantities for the 8 cities plotted versus
the area under the top part of the curves in Figure \ref{fractspeedtime_rush_allcities}.
The quantities considered are: (top left): the average occupied road length per car,
(top right) the average occupied road speed per car (road speed is a function of the
road and is not the car's speed), (bottom left): the road length per occupied road, and
(lower right): the average number of cars per occupied road.  Aside from quantities that
result from congestion, there is no evidence for a cause of congestion in properties of
the roads themselves, leading to the inference that the cause begins at the intersections.}
\label{speedlength_80_500dots}\end{figure}

Figure \ref{speedlength_80_500dots} shows only weak correlations between these four
quantities and the degree of gridlock. An obvious relation is in the lower right,
where the number of occupied roads per car drops for the worst gridlock cases. This
merely reflects the inability of cars to reach their destinations in these cities,
and is more a result of gridlock than a determinant. In the upper right, the average
road speed per car drops for the red point, which is Nairobi; this is not the car
speed but the road speed limit. Still, even with slow road speeds, a high fraction
of the cars are moving more than 30\% slower.

The road length per occupied road and the road length per car seem to increase with
gridlock. This was not expected because it means there is more room in the road
network for cars when gridlock is worse. We would have expected the opposite, that
gridlock results when there is less room on the road for the existing cars.

These considerations lead us to suspect that the intersections are more the problem
than the roads. To study this, we plot on the left in Figure
\ref{speedlength_80_500_top2} the number of roads with slow cars (defined as above
as cars moving at less than 30\% of their local road speed) versus the fraction of
the occupied roads with slow cars. Time varies clockwise around the jagged curves.
The bottom panel has a bigger scale than the top panel so that all of the cities can
be seen clearly in one panel or another. Similarly, the right-hand side of Figure 7
shows the number of roads with slow cars versus the average fraction of the road
speed for all cars.

\begin{figure}
\includegraphics[width=.8\textwidth]{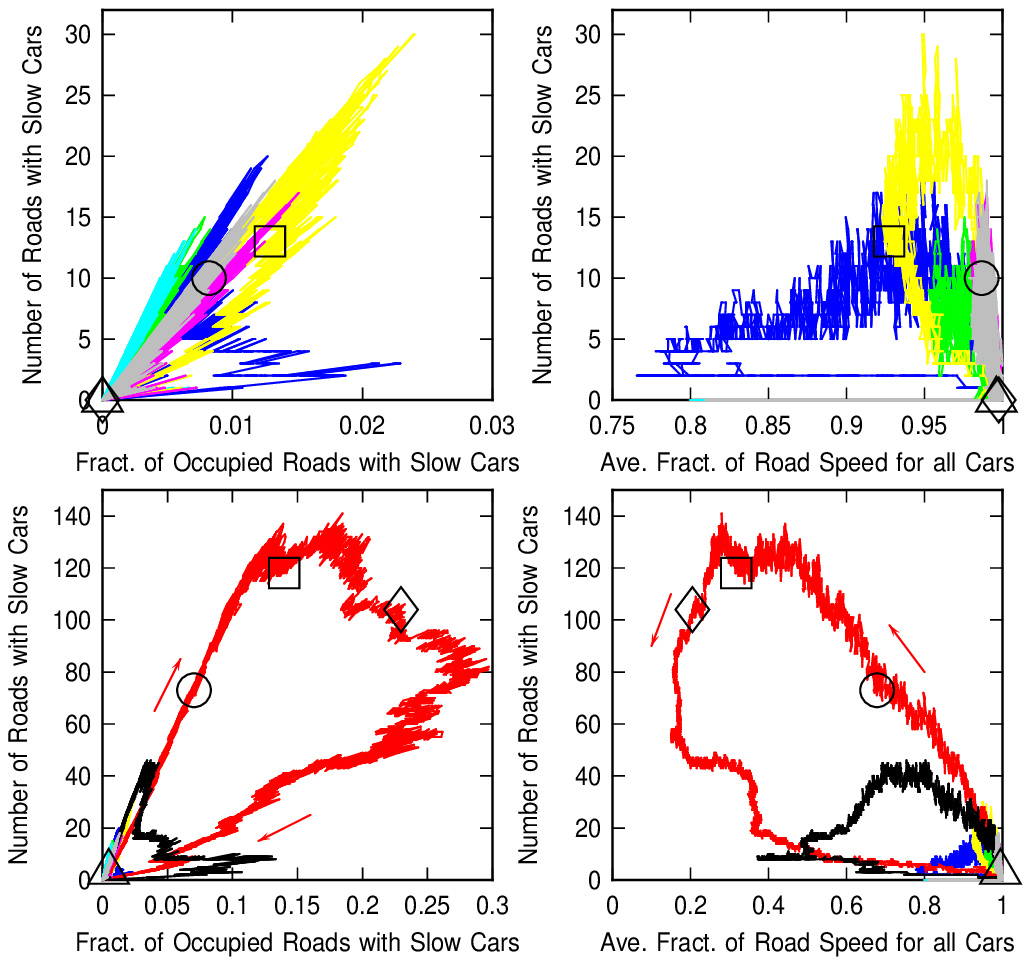}
\caption{(Left:) The number of roads with slow cars (defined to have less than
30\% of the road speed) versus the fraction of the occupied roads with
slow cars. (Right:) The same quantity versus the average fraction of the
road speed for all cars. Each curve is for a different city during the rush hour
models shown in Figure \ref{fractspeedtime_rush_allcities} using the same color
scheme. Points
move clockwise around the curves in time for the left-hand panels and
counter clockwise for the right-hand panels. The top two panels show
enlargements of the lower two panels and also exclude
the red and black curves for clarity. These figures indicate that
gridlock is dominated by a few intersections in a city.}
\label{speedlength_80_500_top2}\end{figure}

These distributions have an interesting pattern. In the left-hand panels, all of the
cities start moving along a diagonal line toward the upper right until about the
time of the peak launch rate, which is shown by the square (symbols are the same as
in Figures \ref{fractspeedtime_rush} and \ref{fractspeedtime_rush_varywidth}). This
trend corresponds to a simultaneous increase in both the number and the fraction of
roads with slow cars. After the time of peak launch rate, the fraction of occupied
roads with slow cars continues to increase as the roads without slow cars begin to
free up.  The roads with slow cars free up much more slowly, decreasing the curve
gradually along the ordinate as it continues to move to the right. Eventually all of
the gridlocked roads begin to empty and the curves decrease down and to the left.
This pattern is another manifestation of the hysteresis seen in Figures
\ref{fractspeedtime_rush}-\ref{fractspeedtime_rush_allcities}, but it shows that
even with bad gridlock only a small fraction of the occupied roads, less than
1\%-10\% for 7 of the cities, actually have this gridlock -- the rest are relatively
free. Also, as shown in the right-hand panels, the average speed of all the cars is
within 80\% of the local road speed for 6 of the 8 cities.  In the worst cases, it
drops down to 20\%.

For Istanbul (black curve), the fraction of occupied roads with slow cars in Figure
7 is a maximum, and the average fraction of the road speed is a minimum, when the
number of roads with slow cars is far lower than the peak number. The same is true
for Washington DC. What this means is that after a while, most of the gridlock is on
only a few roads and the rest of the roads are relatively clear. It takes a long
time for these blocked roads to free up while all the other roads empty. Cities
without this pattern, such as Damascus (yellow) and Beijing (gray) have
distributions that go up and come down on nearly the same diagonal line. For these
cases, the roads that block up easily also free up easily. Thus the openness of the
curves in Figure 7 indicates the range in the ability of blocked roads to free up.
Narrow curves have a small range, which means the troublesome roads and
intersections are all about the same, while open curves have a wide range, which
means that some intersections are much worse than others.

At this point, one of the limitations of the {\it Megaffic} code should be recalled,
as the results for real cities could be different from what we simulate. This
limitation is that the routes used by all of the cars are determined and fixed
before the simulation begins, so drivers cannot change their routes as the
congestion develops. In one sense this is realistic because drivers sometimes have
few options for different routes and are forced to follow the congested roads.
However, some cities have many more side streets connected to the main thoroughfares
than other cities do, and for these well-connected cities, smart drivers will get
off the congested roads and take alternate routes even if they are longer.

We experimented with more diverse route plannings on a square grid of roads to see
what possible improvements there might be. The road grid measured 51 by 51 roads
intersecting at right angles; it will be discussed later for another purpose in
Section \ref{ideal}.  We used Dijkstra's (1959) algorithm to design trips. In one
case, we generated 100 trips that start from the upper-right cross-point in the
square grid to the lower left cross-point using Dijkstra's shortest path for all of
them. As a result, they followed the same path. In a second case, we generated 100
trips from the same start and end points and each trip again used the shortest path
algorithm but the cost of the used roads was raised sequentially as the trips were
generated. The result was a sequence of trips distributed all over the map. The
second approach improved the average trip time in this idealized model by a few per
cent and avoided jams at crowded intersections.

Another experiment moved cars from the entire left-hand column to the entire
right-hand column in a square $51\times51$ road grid. Dijkstra's method decreased
the total trip time by 7\% compared to the {\it Megaffic} algorithm for trip
designs.

These tests suggest that modest levels of improvement are possible with trip designs
that program in some avoidance strategy.  Square grids are optimum for this however,
because the number of routes with the same total road length is enormous. Small
variations in routing can decrease the traffic flow on each road by the inverse of
the number of different routes.

\section{Idealized Road Networks}
\label{ideal}

Idealized road networks allow us to study the most basic properties of traffic flow
without the complexity that comes from a mixture of road structures in real cities.
We considered the three idealized networks shown in Figure \ref{megaffic_roadmake5};
only the inner portions of the left and middle networks are shown.  On the left is a
square grid composed of 51 single lane roads in each direction; the dots are the
intersections. Each road segment has the same length $L$ and the same speed  $v$. In
the center is a triangular network with single-lane roads in each direction along
the $60^\circ$ angles, and with segment lengths $L$ and uniform speeds $v$. On the
right is a square grid with two-way, single-lane segments as before, but now the
central parts along each axis have roads 10 times longer. This third case is
intended to simulate cities with boulevards or highways that can hold more cars than
the shorter side streets elsewhere. In another set of simulations we considered
higher or lower road speeds for the same middle segments in the vertical and
horizontal directions, but now with normal short road lengths there as in the left
panel.

\begin{figure}
\includegraphics[width=1.\textwidth]{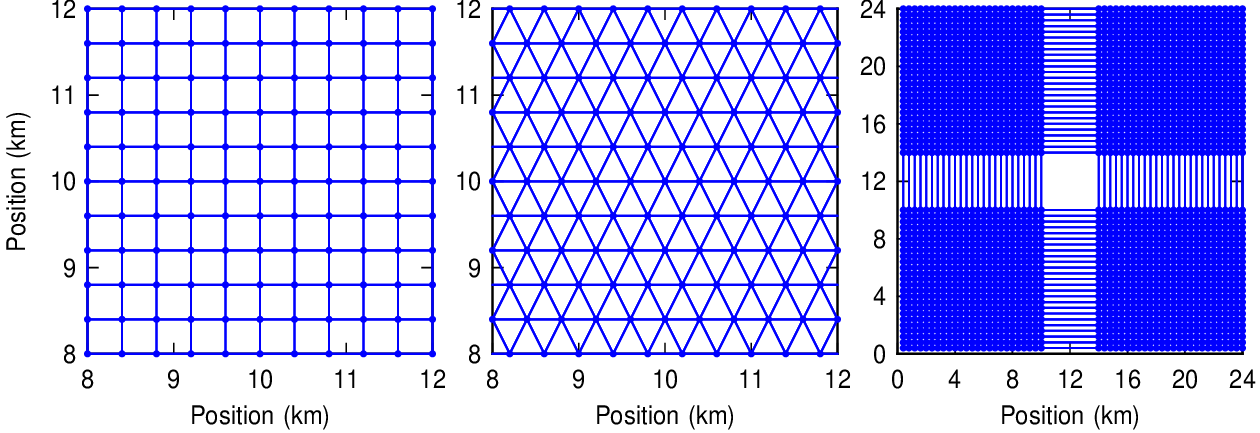}
\caption{Idealized road grids: (left) The inner portion of a square grid
with a total of 51 vertical and 51 horizontal roads. Each segment is 400 m long
and consists of single-lane
traffic in each direction. The axes labels give the road positions in km.
(middle) The inner portion of a triangular
grid with 51 roads in
each $60^\circ$ direction supporting single-lane two-way traffic; each segment is
400 m long. (right) The full image of a square grid
with 51 roads in each direction and roads 10 times longer than the others
in the central cross; the roads are also single-lane and support traffic
in each direction. Segment lengths are 400 m in each quadrant corner and
4 km in the central cross.} \label{megaffic_roadmake5}\end{figure}

The results of rush hour launch rates for these cases are shown in Figure
\ref{square}. Each panel has five cases for the grid type indicated: four with
$(R_0,t_{\rm max})=(80,5000)$ and $(160,5000)$ for each of $(L,v)=(200,30)$ and
$(400,60)$, and a fifth with $(R_0,t_{\rm max})=(40,10000)$ and $(L,v)=(400,30)$.
Units of $R_0$ are cars per 10 seconds; units of $t_{\rm max}$ are seconds; units of
$L$ are meters, and units of $v$ are km hr$^{-2}$.  This choice of cases is made
because cases with $(L,v)=(200,30)$ and $(400,60)$ have the same average travel
times (i.e., from the ratio of road length to speed), which normalizes the
simulations to time. The different $(R_0,t_{\rm max})$ give light and heavy rush
hour traffic with one having twice the launch rate as the other. The fifth case has
half the launch rate for the same number of cars compared to $(R_0,t_{\rm
max})=(80,5000)$, but the road density is the same because the speed is half
compared to the case $(L,v)=(400,60)$. For comparison, the square grid results from
the top left are repeated in the lower left as dotted curves.

\begin{figure}
\includegraphics[width=1.\textwidth]{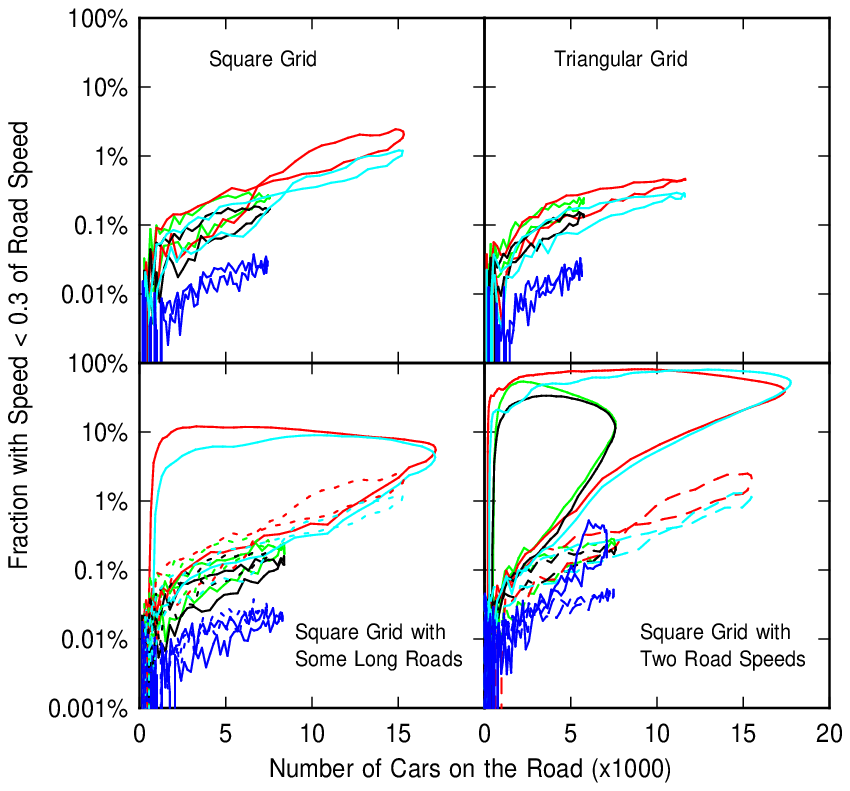}
\caption{The fraction of cars moving slower than 0.3 of the local road speed
is plotted versus the current number of cars on the road for the
idealized grids shown in Figure \ref{megaffic_roadmake5}. Each panel has
5 rush hour launch rates characterized by the parameter combinations
$(R_0,t_{\rm max},v,L)=(80,5000,60,400)$ for green curves,
$(160,5000,60,400)$ for red, $(80,5000,30,200)$ for black,
$(160,5000,30,200)$ for cyan, and $(40,10000,30,400)$ for blue. The
meaning and units of these parameters are, in order, peak launch rate in
cars/10s, total time span for car launching in seconds,
road speed in km hr$^{-1}$, and segment length in m. The top left
and right panels show only these curves. The lower left panel shows
solid-line curves for the square grid with long roads in a cross pattern (Fig.
\ref{megaffic_roadmake5}), and it repeats the curves in the upper left as
dotted lines, for comparison. The lower right panel is for a uniform square
grid like the upper left panel, but the road speed in the segments of the
central cross is 2 times higher (solid curves) or 2 times lower (dashed
curves) than the speed in the other roads. Road networks with long segments (lower left) or
road networks with fast segments (lower right) experience greater congestion at high
launch rates because these enhanced road segments can have a lot of cars but they have
only one endpoint to exit onto the rest of the grid.
} \label{square}\end{figure}

The results show relatively little gridlock for the square and triangular grid cases
(top left and right) unless there are long roads mixed with short roads (lower
left). Then the congestion gets much worse for the heavy rush hour cases (red and
cyan curves in the lower left). This worsening condition contrasts with the light
rush hour case in the lower left, where long roads improve the flow (blue, black and
green solid-line curves have slightly smaller slow-car fractions than the dotted
curves of the same colors). For equal road lengths, the triangular grid is
marginally better than the square grid. In all cases, the lowest launch rates (blue
curves) have the least congestion.

The results for a square grid with uniform road lengths and variable road speeds are
shown in the lower right of Figure \ref{square}. As mentioned above, all of the road
speeds are the same except for horizontal roads in a vertical strip through the
center and vertical roads in a horizontal strip through the center, where the roads
are either half the speed of the other roads (dashed curves) or twice the speed
(solid curves).  Lowering the speed of some fraction of the roads does not increase
congestion noticeably (the dashed curves in the lower right panel are like the
similarly-colored curves in the upper left).  However, increasing some road speeds
creates problems for all launch rates. The reason for this is that cars on the fast
roads come to their ending intersections quickly, and then they have to wait for the
cars ahead of them to cross before they can go.

Note that the nominal road speeds in all of these cases can be reached after
traveling only at most 20\% of the road length, so the congestion is not an artifact
of acceleration in a limited domain. For example, several cases in Figure
\ref{square} have a road length of 200 meters with a road speed of 30 km hr$^{-1}$,
which is 8.3 m s$^{-1}$. The acceleration is always $1.7\pm0.3$ m s$^{-2}$. With
this acceleration, the road speed is reached after traveling only 20 m, which is
10\% of the road length. For 60 km hr$^{-1}$ roads of 400 m length, the speed is
reached after 20\% of the road is traveled.

\section{Dependence on Acceleration}
\label{accel}

Merging and leaving an intersection or other stopping point should be faster if the
acceleration is higher. The dimensionless quantity $aD/v^2$ was discussed in the
introduction. For a given road density, written here as the separation between cars,
$D$, and road speed, $v$, a higher acceleration $a$, and higher corresponding
braking rate, which in our simulations is proportional to $a$, allow for easier
merging between cars, either during lane shifts or while entering new roads at
intersections.  To test for this in {\it Megaffic}, we increased the mean and
standard deviation for the accelerations of all cars by factors of $2^{0.25i}$ for
$i=0,$ 1, 2, up to 9. Thus the acceleration ranges between the nominal value we have
been using in Figures 1-7, which is $a=1.7\pm0.3$ m s$^{-2}$ up to $a=8.09\pm1.43$ m
s$^{-2}$.

The rush hour models with these accelerations are shown in Figure
\ref{fractspeedtime_rush_varyacc} for Washington DC and Damascus.  The looping
curves decrease rapidly with increasing acceleration. After reaching a certain
value, they increase again but only at a level of 0.1\% or so. Too large an
acceleration increases the gridlock because then cars move too fast after a single
time step to merge with traffic at the nominal road speed.

\begin{figure}
\includegraphics[width=1.\textwidth]{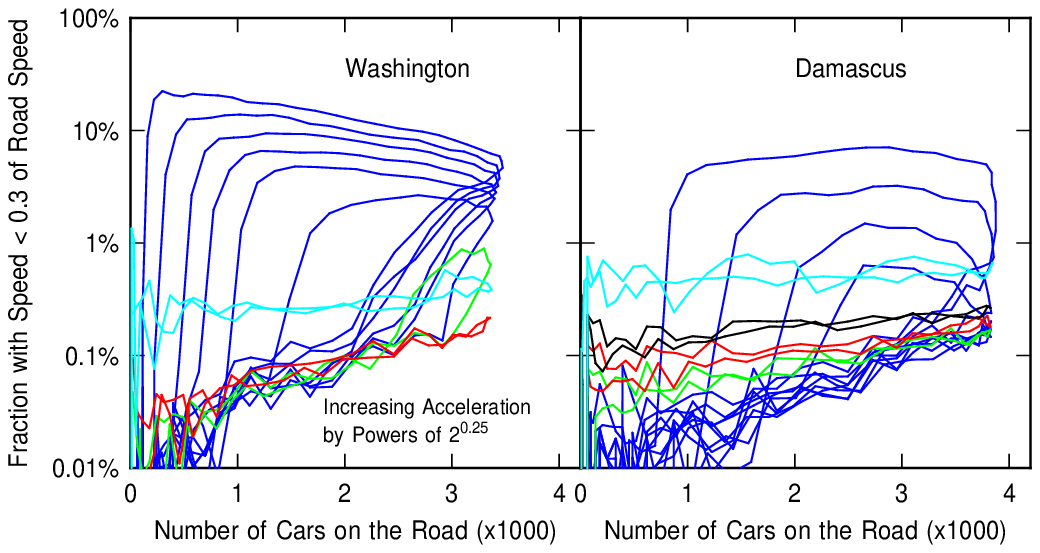}
\caption{The fraction of slow cars versus the number of cars for
two cities, Washington DC and Damascus, with $(R_0,t_{\rm max})=(80\;{\rm cars/10s},5000\;{\rm s})$
and different accelerations, increasing by powers of $2^{0.25}$ from the top
curve down to the green curve and then back up again to the cyan curve.
Greater average car acceleration decreases congestion. } \label{fractspeedtime_rush_varyacc}\end{figure}

Figure \ref{int_vs_acc} shows the decrease in area under the hysteresis loop versus
the acceleration factor for the five cities in Figure
\ref{fractspeedtime_rush_allcities} that have the most gridlock. Each city improves
when the acceleration increases, with the least gridlocked cities improving the
fastest. Highly gridlocked cities do not improve as much with acceleration because
each bad intersection has a lot of stopped cars and only a small fraction of cars
get to accelerate at the beginning of the queue when it leaves the intersection.
Improvements from increased acceleration help more when there are small queues at a
large number of intersections, rather than large queues at a few intersections.

\begin{figure}
\includegraphics[width=.8\textwidth]{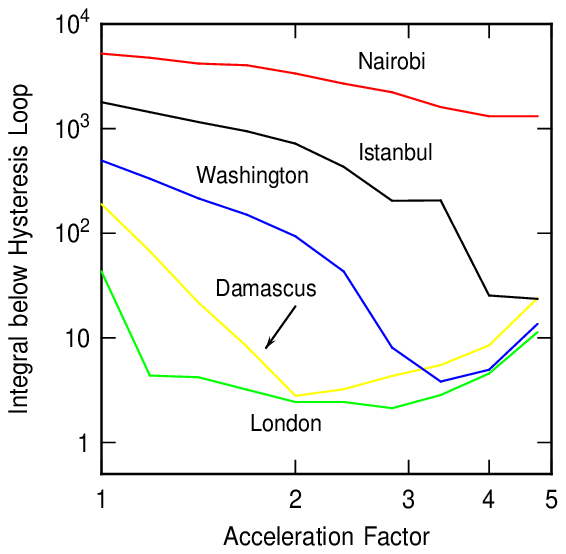}
\caption{The integral below the curve of slow car fraction versus
car number -- the hysteresis loop -- is shown versus the acceleration
factor for simulations of 5 cities. The acceleration for each point
in a segmented curve equals the nominal acceleration, $1.7\pm0.3$ m s$^{-2}$
multiplied by the acceleration factor. Highly congestible cities like
Nairobi and Istanbul are less sensitive to acceleration than weakly
congestible cities because the congestible cities have a smaller
number of more difficult intersections where the queue to cross is long.
Then only a small fraction of slow cars
get a chance to accelerate up to the road speed while the rest of the
cars wait in the queue.} \label{int_vs_acc}\end{figure}

\section{Results for Steady Flow Simulations}
\label{steady}

Steady traffic flow inside a city center was also investigated using a vehicle
launch rate that increases first as a half-Gaussian with $\sigma_{\rm a}=1000$
seconds up to the peak rate $R_0$ at $t_0=2500$, as for most of the rush hour
simulations discussed above, and then levels off to the steady rate $R_0$ for
another 20,000 seconds.  At low $R_0$, traffic was stable with the rate of trip
completion equaling the launch rate. As $R_0$ increased, there was a certain value
beyond which the number of cars on the road increased indefinitely, causing more and
more congestion over time.

Figure \ref{megaffic_fractspeedtime_ramp4} shows the results for one city; other
cities are similar. The launch rates increase from $R_0=10$ cars in each 10 second
interval, up to 100 cars/10 s, in steps of 10 cars/10 s. In the lower right panel,
the number of cars is shown for each $R_0$ as a function of time using logarithmic
coordinates on the ordinate. The lower curves level off, indicating a constant
number of cars or equilibrium between trip starts and completions. Higher launch
rates have increasing numbers of cars without leveling off. The center panel on the
right plots the same thing but in linear coordinates on the ordinate, to emphasize
the rapid increase in car counts for large $R_0$ at later times.  The lower and
middle left panels show the summed speeds of all the cars versus the number of cars
and the time, respectively. The summed speed is a measure of the total traffic flux.
Higher $R_0$ gives higher summed speed even at late times and high car numbers, so
the city is supporting these cars and still moving them. However, the average speed
per car, shown by the red decreasing curves in the bottom left, decreases rapidly
with increasing car numbers, suggesting congestion. This congestion is shown better
in the top panels where the fraction of cars with speeds less than 30\% of their
local road speed is plotted versus time and number of cars on the road. This
fraction stays low for low $R_0$ but increases to a saturated value of $\sim0.9$ at
high $R_0$ and late times. It does not reach unity because even with congestion,
there are still cars that move on unblocked routes.

\begin{figure}
\includegraphics[width=.75\textwidth]{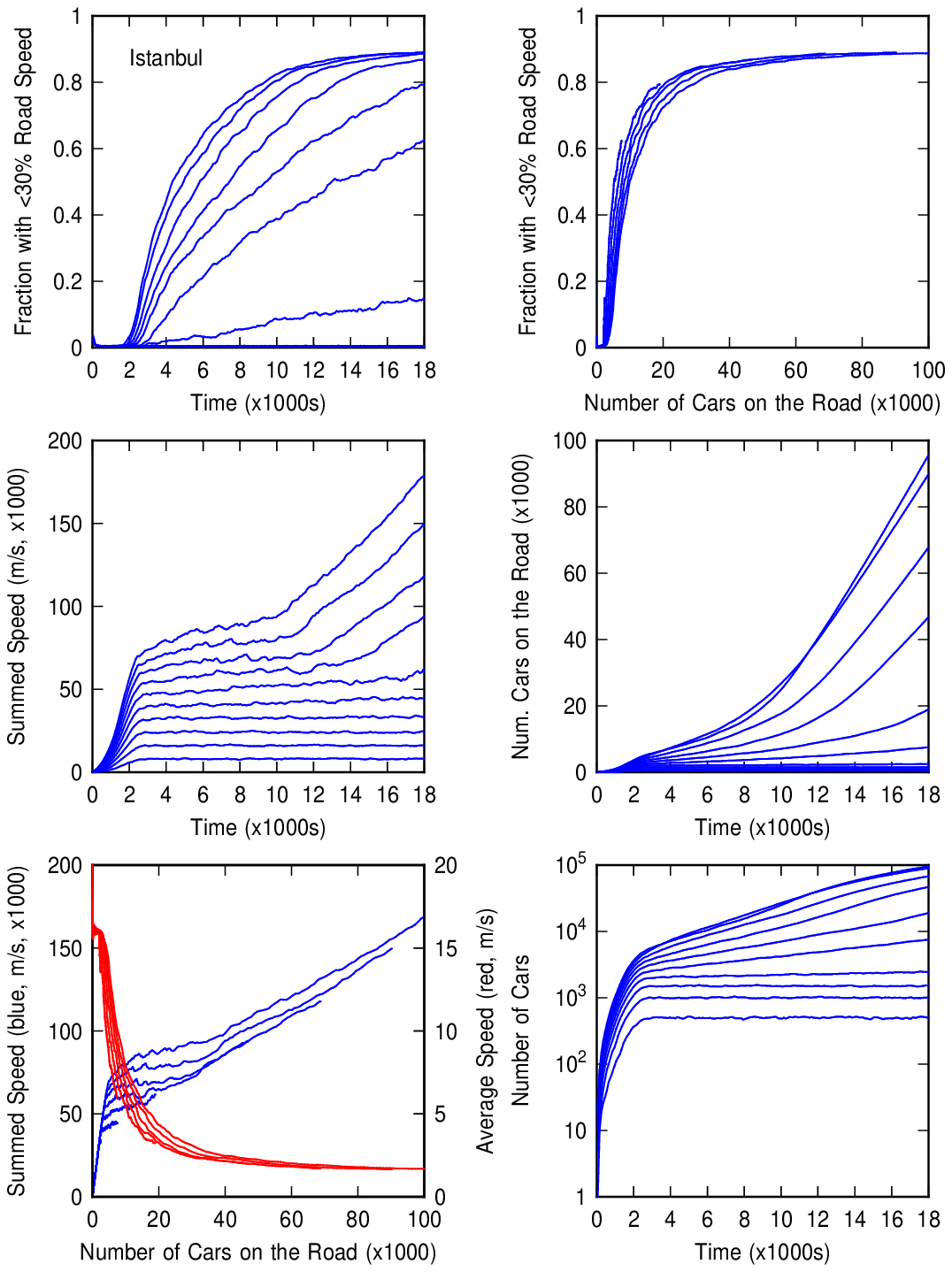}
\caption{These curves show the time development of traffic in one representative city,
Istanbul, when the rate of car launching increases and then levels off to
a value $R_0=10$, 20, ... 100 cars per 10 seconds. For small launch rates, the
number of cars on the road can remain constant because the starting and ending
rates are equal. For large launch rates, the number of cars continuously
increases and the average speed per car continuously decreases (red curves, lower
left) because of increasing congestion. The various panels are described
in the text. Note that the summed speed is a measure of the total car flux
in and around the city, and it continuously increases with time (middle left panel)
even as the congestion increases (top left panel).
The relationship between the fraction of slow cars and the number of cars
on the road is nearly independent of the launch rate in a steady state (upper right).}
\label{megaffic_fractspeedtime_ramp4}\end{figure}

The maximum $R_0$ for equilibrium flow varies for the 8 cities in the same way that
the congestion indicator varies for the rush hour experiments. This variation is
shown in Figure \ref{versuslaunchrate} where the number of cars on the road at
10,000 seconds is plotted versus the launch rate $R_0$ on log-log axes. The 4 cities
on the left have the largest congestion and are plotted separately from the 4 cities
on the right, using a different range of coordinate values. Following the order of
cities in Figure \ref{fractspeedtime_rush_allcities}, those that are most easily
congested in a steady state branch off earlier in Figure \ref{versuslaunchrate} from
the linear increase of car numbers with $R_0$.

\begin{figure}
\includegraphics[width=1.\textwidth]{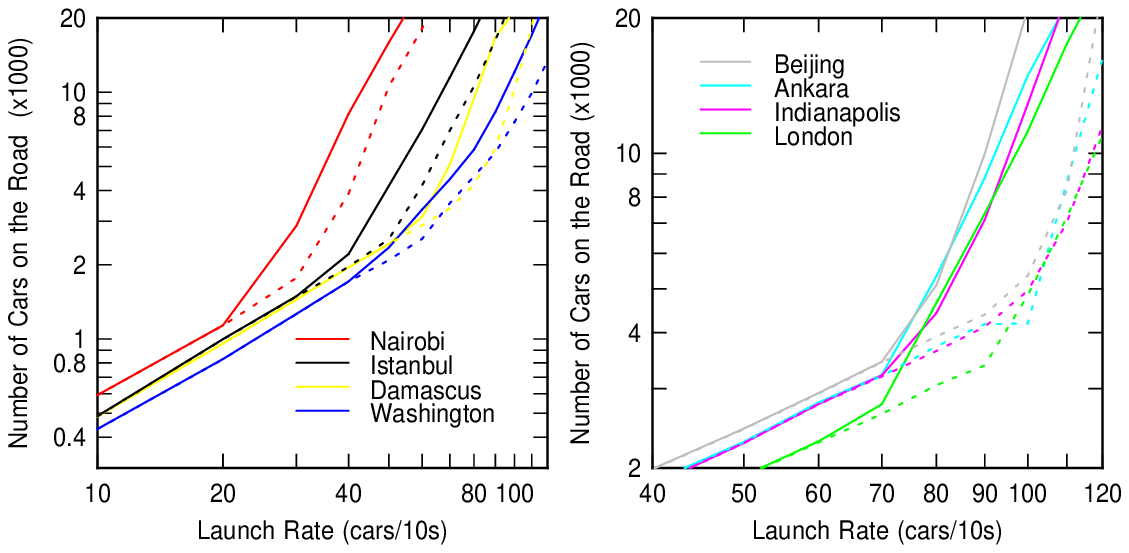}
\caption{The number of cars on the road versus the launch rate for the
case of steady flows shown in Figure \ref{megaffic_fractspeedtime_ramp4}.
The number increases linearly with launch rate for small launch rates
when the roads are uncongested, but the number increases much
faster when the launch rate reaches a critical value and the roads begin fill up.
The critical launch rate increases with city in the same order as the hysteresis
loops decrease in Figure \ref{fractspeedtime_rush_allcities}.
} \label{versuslaunchrate}\end{figure}

The solid curves in Figure \ref{versuslaunchrate} are for the normal acceleration,
$a=1.7\pm0.3$, used above, and the dotted curves are for twice this acceleration
with the same road speeds. As in Section \ref{accel}, increased acceleration
decreases congestion. In this case, the decreased congestion allows higher launch
rates and more cars on the road in a steady state before the runaway growth begins
at high $R_0$.  Thus the dotted lines lie below and to the right of the solid lines.
The result is sensitive to acceleration as found above: launch rates can increase
safely by $\sim25$\% if the average acceleration doubles.

\section{Conclusions}
\label{conclusions}

City traffic was investigated using the IBM software package {\it Megaffic} combined
with car-by-car and second-resolved streaming analytics using socket writes to a
second computer. Rush hour and sustained traffic flows in 8 cities were followed
with Gaussian-shaped vehicle launch rates and random city routes determined in
advance.

A good measure of congestion was found to be the fraction of cars moving slower than
30\% of their local road speed. Decreasing the launch rates for the same window of
time, or increasing the time interval for vehicle launching with the same total
number of cars, both decreased congestion, as expected from common experience.
Increasing vehicle acceleration for the same road speed also improved traffic flow
as it increased the probability that a car waiting at an intersection could enter
the next road and merge safely.

The main impediments to traffic flow seemed to occur at the intersections in our
models, not in the free-streaming traffic between intersections. Cars stopped at an
intersection had to wait for an opening to cross to the next road, and all of the
cars behind them had to wait also. Increasing car acceleration helped, as just
mentioned, but only for the lead car at the intersection. If a road system was
jammed to a certain level by a small number of cars at each of a large number of
intersections, then increased acceleration improved the overall flow rate, sometimes
by a large factor. However, if the same level of jamming was caused by a large
number of cars stuck at a small number of intersections, then vehicle acceleration
did not matter much as the fraction of cars with improved mobility was small.

Real-time streaming analytics using all of the data generated by {\it Megaffic} was
found to help in visualizing and understanding problems as they arose. It would have
been impossible with the available computers to store all of the data and analyze it
later in large-network simulations. Standard storage and retrieval methods used for
limited runs could not be scaled to real-life systems.

{\it Acknowledgements}: We are grateful to the IBM Mega Traffic Simulator team for
providing their software for our modification and use; key assistance came from
Sachiko Yoshihama, Hideyuki Mizuta, Kumiko Maeda, Toyotaro Suzumura, Takayuki
Osogami, Tsuyoshi Ide, and particularly Takashi Imamichi, who answered our weekly
questions. We are also grateful to Alain Biem for guidance on streaming analytics
throughout this project and on programming tips for {\it Infosphere Streams}.
Additional support and suggestions from Supratik Guha are much appreciated.

\end{document}